\documentclass[pra,twocolumn,a4paper,showpacs,superscriptaddress]{revtex4}
\usepackage{amsmath}
\usepackage{amsfonts}
\usepackage{graphicx}

\begin{document}

\title{Entanglement gauge and the non-Abelian geometric phase with
  two photonic qubits}
\author{Karl-Peter Marzlin}
\affiliation{Department of Physics and Centre for Advanced Computing
  -- Algorithms and Cryptography, Macquarie University, Sydney, New
  South Wales 2109, Australia}
\affiliation{Fachbereich Physik der Universit\"at Konstanz, Postfach
  5560 M674, D-78457 Konstanz, Germany}
\author{Stephen D. Bartlett}
\author{Barry C. Sanders} 
\affiliation{Department of Physics and Centre for Advanced Computing
  -- Algorithms and Cryptography, Macquarie University, Sydney, New
  South Wales 2109, Australia}
\date{2 March 2003}

\begin{abstract}
  We introduce the \emph{entanglement gauge} describing the combined
  effects of local operations and nonlocal unitary transformations on
  bipartite quantum systems.  The entanglement gauge exploits the
  invariance of nonlocal properties for bipartite systems under local
  (gauge) transformations.  This new formalism yields observable
  effects arising from the gauge geometry of the bipartite system.  In
  particular, we propose a non-Abelian gauge theory realized via two
  separated spatial modes of the quantized electromagnetic field
  manipulated by linear optics.  In this linear optical realization, a
  bi-partite state of two separated spatial modes can acquire a
  non-Abelian geometric phase.
\end{abstract}
\maketitle

\section{Introduction}

At the heart of quantum information theory is the phenomenon of
entanglement between spatially separated systems.  Nonlocal
correlations due to entanglement allow powerful information processes
that cannot be performed classically: entanglement is central to tests
of local realism~\cite{Bel64} and is a fundamental resource of quantum
teleportation~\cite{Ben93}, quantum
cryptography~\cite{Eke91,Cir97,Fuc97}, and possibly quantum
computing~\cite{Bra99,Eke98}.  Moreover, nonlocal
entanglement-generating transformations of multi-partite states can
allow for communication~\cite{Ben92,Eis00} and distributed quantum
computing~\cite{Eis00,Cir99}.  The possibilities of manipulating
nonlocal properties using only local operations (e.g., entanglement
distillation~\cite{Ben96}) are therefore an important resource issue
for quantum information.

Here we show that nonlocal properties of quantum states are simplified
and isolated through the use of gauge theoretical
concepts~\cite{Egu80}.  A powerful tool of modern mathematical
physics, gauge theory is used to describe the geometric structure of
systems possessing symmetry under specific (gauge) transformations.
In quantum mechanics, it is known that local transformations on
components of a multi-partite system do not change the entanglement.
We show that local operations can be expressed as a form of gauge
transformation, which we define as the \emph{entanglement gauge}.
Nonlocal properties such as entanglement, then, are naturally
expressed as entanglement gauge invariant quantities.  With this
entanglement gauge formalism, general transformations on a bi-partite
system can be decomposed into nonlocal and local (gauge)
transformations.  Nonlocal properties can be defined on a generally
curved space that is given by an equivalence class of states under
local operations.

One manifestation of employing entanglement gauge theory is that, for
a given nonlocal transformation on this generally curved space,
effects due to geometric phases~\cite{Ber84,Sim83} can arise in
non-trivial ways.  These phases can arise in many different physical
situations~\cite{Fue02} and may even be utilized as a resource for
quantum information processes; of particular interest, it has recently
been proposed~\cite{Jon00} that geometric phases in two-qubit systems
can allow for a fault-tolerant conditional phase shift gate in quantum
computation.  For general systems with more than two qubits, holonomic
quantum computation has also been investigated~\cite{Zan99}.

Current experiments are now at the point where controlled local and
nonlocal transformations in a wide variety of physical systems are
accessible, and one can observe the manifestations of these geometric
phases.  Optical realizations offer the advantage of negligible
decoherence as well as the advanced ability to implement unitary
operations using linear optics.  These realizations also provide a
natural source of entanglement in parametric down-conversion (PDC),
with which photon pairs can be created in a polarization-entangled
state~\cite{Kwi95}.  Recent experiments~\cite{Whi02} have produced a
wide selection of two-photon states, with varying degrees of
entanglement and disorder (entropy), and have characterized the
resulting state using quantum tomography.

Many optical experiments can be described as first producing the
photon pairs via PDC (with controllable degrees of entanglement and
entropy), and then directing the photons through passive linear
optical elements (beam splitters, phase shifters, polarization
rotators).  Using a setup of this form, Kwiat and Chiao~\cite{Kwi91}
have demonstrated an Abelian geometric phase shift in an (unentangled)
two-photon system.  The Abelian geometric phase for an entangled or
partially-entangled system has been investigated
theoretically~\cite{Sjo00a}, and an Abelian geometric phase for mixed
states in interferometry has been proposed~\cite{Sjo00b}.

We show in this paper that optical states of a bi-partite system are
naturally described in an entanglement gauge formalism.  Also, by
manipulating such states with linear optical elements, they can
acquire a non-Abelian geometric phase (NAGP)~\cite{Wil84,Ana88}.  An
NAGP arises if instead of a single state vector, which spans a
one-dimensional subspace of Hilbert space, the cyclic evolution of a
$n$-dimensional subspace is studied.  In this case, the usual U(1)
geometric phase factor is generalized to geometric U($n$) unitary
transformations.  The notion of a non-Abelian ``phase'' is justified
because all eigenvalues of unitary operators are phase factors,
despite the fact that the resulting transformation on a general state
is not simply an additional phase in the traditional sense.  We
illustrate the concept of a NAGP using quantum interferometry to
evolve bi-partite states about a closed loop; the NAGP acquired can be
measured using quantum tomography, a technique that can completely
characterize our proposed bi-partite states~\cite{Whi02}.  We discuss
its relevance in the context of entanglement and quantum information
theory.

The paper is structured as follows.  In Sec.~\ref{sec:Interferometry},
we describe the relevant Hilbert spaces and transformations in quantum
interferometry, along with a description of the entanglement gauge
structure.  We define the NAGP for cyclic evolution in
Sec.~\ref{sec:Gauge}, and give a parametrization and explicit
expression for the NAGP for the relevant coset space.  In
Sec.~\ref{sec:NAGP}, we calculate the NAGP acquired in a quantum
interferometry setup for various types of cyclic evolution, giving
specific examples.  We conclude with Sec.~\ref{sec:Conc}.

\section{Photon interferometry}
\label{sec:Interferometry}

In this section, we review the mathematical structure of two-photon
quantum interferometry.  This structure allows us to construct the
geometrical space describing the entanglement gauge for bi-partite
optical states of two photons.  (We note that the general formalism
developed here can be applied to other physical systems, e.g., trapped
ions~\cite{Cir95} with two phonons.  However, working explicitly with
an optical realization gives a valuable physical context.)

Consider a two-channel (four-port) optical interferometer with
polarization-dependent elements.  There are four boson field modes to
consider, each with a corresponding annihilation operator: $a_H$
corresponding to the horizontal polarization for the $a$ spatial mode,
$a_V$ corresponding to the vertical polarization for the $a$
spatial mode, and annihilation operators~$b_H,b_V$ for the
horizontal and vertical polarizations for the $b$~spatial modes.  A
passive linear optical experiment can employ polarization rotation,
beam splitters, phase shifters and mirrors as stages of the processing
of the quantum state~\cite{Bar01}.  Each of these stages can be
represented mathematically as a unitary transformation provided that
losses are neglected.  Together, these transformations close to the
group $U(4)$; thus, we say that a passive polarization-dependent two
channel interferometer invokes a transformation $g \in U(4)$.

For a quantum interferometer, transformations on quantum optical
states are given by a representation of $U(4)$.  Because passive
optical transformations are photon-number-preserving, each irreducible
representation (irrep) is labelled by $N$, the total number of
photons.  That is, the Hilbert space $\mathbb{H}_N$ for each irrep $N$
contains the $U(4)$ highest weight state $|\phi_N\rangle =
\frac{1}{\sqrt{n!}}(a^\dagger_H )^N |0\rangle$, where $|0\rangle$ is
the Fock state vacuum.  This highest weight state is constructed such
that all the photons are in channel $a$ with horizontal polarization.

In the following, we define local operations to be those operations
that act on the spatial modes $a$ (or $b$) alone, whereas nonlocal
operations mix the spatial modes $a$ and $b$ together.  Of course,
nonlocal operations in $U(4)$ are performed using only spatially local
interactions by bringing modes $a$ and $b$ together, such as at a
beamsplitter.  For our development, however, we define these
transformations to be nonlocal.  (One can consider the $U(4)$
interferometer to be a ``black box''; the internal workings may bring
together the spatially-distinct modes $a$ and $b$, but we consider
only the resultant effect on the joint state of these modes.)
Polarization rotations and polarization-dependent phase shifts in
channels $a$ and $b$ describe local operations on mode $a$ and $b$ and
form the subgroups $U(2)^a$ and $U(2)^b$, respectively.  We define $LO
= U(2)^a \times U(2)^b \subset U(4)$ as the subgroup of local
operations.  In contrast to these local operations, the group $U(4)$
also contains transformations such as those describing a beamsplitter,
which interact the two spatial modes $a$ and $b$ and are nonlocal by
our definition.

Define $\mathbb{H}_2$ to be the Hilbert space of all two-channel
polarization-dependent states with exactly two photons.  This Hilbert
space is 10 dimensional, and is the carrier space for the two-photon
($N=2$) irrep of $U(4)$.  The space $\mathbb{H}_2$ is relevant to
quantum optics and quantum information because it includes the space
of possible output states from PDC and in particular the
maximally-entangled Bell states.  It should be noted that the action
of $U(4)$ is \emph{not} transitive on the Hilbert space
$\mathbb{H}_2$: it is not possible to perform arbitrary unitary
transformations on two photons using only linear optics~\cite{Lut99}.

Nondegenerate PDC produces two distinguishable photons, one in each
spatial channel.  The output states of PDC lie in the subspace of
$\mathbb{H}_2$ spanned by the four states
\begin{align}
  \label{eq:NondegeneratePDCSubspace}
  |HH\rangle &= a_H^\dag b_H^\dag|0\rangle \, ,& |HV\rangle &=
  a_H^\dag b_V^\dag|0\rangle \, , \nonumber \\
  |VH\rangle &= a_V^\dag b_H^\dag|0\rangle \, ,& |VV\rangle &=
  a_V^\dag b_V^\dag|0\rangle \, .
\end{align}
This subspace, denoted $\mathbb{H}_{(1,1)}$, is the carrier space for
$(1,0) \times (1,0)$ irrep of the group of local transformation $LO =
U(2)^a \times U(2)^b$; thus, local transformations leave this subspace
invariant.  An alternative basis is given by the Bell states
\begin{align}
  \label{eq:BellStateBasis}
  |\Psi^{\pm} \rangle &= \tfrac{1}{\sqrt{2}} \bigl( |HV\rangle \pm
  |VH\rangle \bigr) \, , \nonumber \\
  |\Phi^{\pm} \rangle &= \tfrac{1}{\sqrt{2}} \bigl( |HH\rangle \pm
  |VV\rangle \bigr) \, .
\end{align}
The subspace $\mathbb{H}_{(1,1)}$ consists of ``qubit states''; each
mode can be considered as a two-level system described by the
polarization state of the photon.  This method for representing qubits
in the polarization state of a photon is known as the dual rail
representation~\cite{Chu95}.  The larger Hilbert space $\mathbb{H}_2$
contains other states (not in $\mathbb{H}_{(1,1)}$) such as the two
photon state $\frac{1}{\sqrt{2}}(a_H^\dag)^2|0\rangle$, which do not
describe ``valid'' qubit states in the dual rail representation.

\begin{figure}
  \includegraphics*[width=3.25in,keepaspectratio]{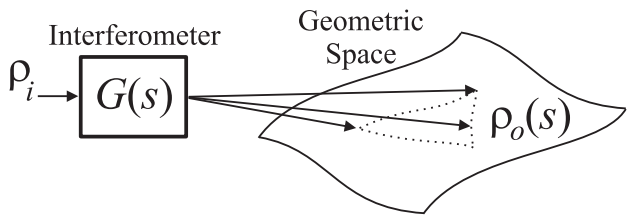}
  \caption{The transformation of the state in the geometric space is
    depicted diagrammatically.  The input state of the interferometer,
    $\rho_i$, is transformed via a parametrized $U(4)$ transformation
    $G(s)$ to an output state $\rho_o(s)$.  By adjusting the
    parameters of the interferometer appropriately, the output state
    can be made to evolve around a closed path in the geometric
    space.}
  \label{fig:GeoSpace}
\end{figure}

We now consider transforming a pure state $\rho$ with support in
$\mathbb{H}_{(1,1)}$ using a linear interferometer; i.e., a $U(4)$
transformation.  Rather than considering the (complex) evolution as
the optical state traverses in time through the elements of the
interferometer, we instead consider transformations in the space of
output states given by adjusting the parameters of the interferometer.
The space of output states of the interferometer, then, is a $U(4)$
orbit of the state $\rho$, given by all the different output states
related to $\rho$ by adjusting the interferometer parameters; see
Fig.~\ref{fig:GeoSpace}.
As discussed in the introduction, we can view local transformations as
gauge transformations that do not alter the entanglement or other
nonlocal properties.  With local transformations described in terms of
this entanglement gauge, we consider two states as equivalent, $\rho'
\simeq \rho$, if $\rho'$ can be obtained from $\rho$ by local
operations only.  With this equivalence, the output of the
interferometer can be identified with the coset space $U(4)/LO$
describing inequivalent states obtained from $\rho$ by nonlocal
operations; we refer to $U(4)/LO$ as the \emph{geometric space}.  A
general parametrized nonlocal $U(4)$ transformation, then, describes a
path $\mathcal{C}$ in this space $U(4)/LO$.  If this path is closed,
the state will return to the same point $\rho' \simeq \rho$ in
$U(4)/LO$, equivalent to the initial state to within a local
operation.  Note that, with this viewpoint, parametrized $U(4)$
transformations in the output space can be implemented without using
time $t$ but instead a \emph{pseudotime} parameter $s$: this
parameter, which is a function of the adjustable parameters of the
interferometer, can be used for \emph{controlled} evolution about
various paths in the output space~\cite{San01,deG01}.

\section{Entanglement gauge structure and the non-Abelian geometric
  phase}
\label{sec:Gauge}

In this section, we show how the gauge structure of this space can
lead to a non-Abelian geometric phase (NAGP)~\cite{Wil84,Ana88} upon
cyclic evolution about a closed path $\mathcal{C}$ in the geometric
space $U(4)/LO$.  A nonlocal transformation is implemented by
adjusting the parameters of the interferometer and is described by the
parametrized transformation $G(s) \in U(4)$, $0 \leq s \leq s_0$.  Let
the $U(4)$ interferometer initially be set to induce the identity
transformation on the input state, so that $G(0)$ is the identity in
$U(4)$. The endpoint is chosen such that $G(s_0)$ closes the path in
the geometric space, i.e., $G(s_0) \in LO$ and thus for any initial
state $\rho$ with support in $\mathbb{H}_{(1,1)}$ the final state is
\begin{align}
  \label{eq:CyclicCondOnG}
  \rho' &= G(s_0)\rho G(s_0)^\dag \nonumber \\ &\simeq \rho \, .
\end{align} 
That is, after cyclic evolution, the transformed state $\rho'$ again
has support in $\mathbb{H}_{(1,1)}$, and is related to the initial
state $\rho$ by a local transformation.  The output state of the
interferometer will follow a closed path $\mathcal{C}$ in the coset
space $U(4)/LO$ parametrized by $s$.  Equivalently, we can think of
the parametrized $U(4)$ transformation $G(s)$ propagating the subspace
$\mathbb{H}_{(1,1)}$ about a closed loop in the Hilbert space
$\mathbb{H}_2$.

Let $|\psi_a(0)\rangle,\ a=1,\ldots,4$ be a basis for
$\mathbb{H}_{(1,1)}$ (e.g., the basis of
Eq.~(\ref{eq:NondegeneratePDCSubspace})).  We can define a transformed
set at each point $s$ along the path as
\begin{equation} 
  |\psi_a(s)\rangle = G(s) |\psi_a(0)\rangle \, .
\end{equation} 
For a closed path $\mathcal{C}$ in the geometric space $U(4)/LO$, the
NAGP is given~\cite{Wil84,Ana88} by the Wilson loop
\begin{equation}
  \label{eq:WilsonLoop} 
  K_{\rm NAGP} = {\rm P} \exp \{ {\rm i} \oint_{\mathcal{C}}
  \mathcal{A}\} \, ,
\end{equation} 
where the gauge potential $\mathcal{A}$ is given in this basis as a
function of the parameter $s$ by
\begin{align} 
  \mathcal{A}_{ab}(s) &= {\rm i} \langle \psi_a(s)| \frac{{\rm d}}{{\rm
      d}s}|\psi_b(s) \rangle\, {\rm d}s \nonumber \\
  & = {\rm i} \langle \psi_a(0)|G^\dagger(s) \frac{{\rm d}G}{{\rm
      d}s}|\psi_b(0) \rangle \, {\rm d}s \, .
  \label{AinG}
\end{align} 
This gauge potential can be expressed in a parameter-independent way
as
\begin{equation}
  \label{eq:GaugePotentialAsMaurerCartan}
  \mathcal{A}_{ab} = \langle \psi_a(0)| \Theta_G |\psi_b(0) \rangle \,,
\end{equation}
where the Lie algebra-valued 1-form $\Theta_G = {\rm i}
G^\dagger {\rm d}G$ is known as the \emph{Maurer-Cartan form}.  In the
following, we will use this Maurer-Cartan form, along with a suitable
parametrization of $G$, to derive an explicit expression for the gauge
potential $\mathcal{A}$.

\subsection{Entanglement gauge transformations}

In this formulation, a local unitary transformation $K(s) \in LO$
corresponds to a gauge transformation.  Restricted to
$\mathbb{H}_{(1,1)}$, it can be viewed as a basis transformation
$|\psi_a^\prime (s)\rangle = K(s) |\psi_a (s)\rangle$ with
$|\psi_a'(s)\rangle \in \mathbb{H}_{(1,1)}$.  One example of a gauge
transformation is a polarization rotation (about any angle) in one
spatial mode.  Under a gauge transformation, the gauge potential
transforms as
\begin{equation} 
  \mathcal{A}^\prime = K^\dagger \mathcal{A} K + {\rm i} K^\dagger
  {\rm d}K \, ,
\end{equation} 
and the Wilson loop (\ref{eq:WilsonLoop}) transforms
covariantly~\cite{Wil84}.

For the special case when $G(s) = K(s) \in LO$, the gauge potential
corresponds to a pure gauge.  In this situation, $K$ restricted to
$\mathbb{H}_{(1,1)}$ satisfies
\begin{align} 
  {\rm d}(K^\dagger {\rm d}K) &= {\rm d}K^\dagger  \wedge {\rm d}K
  \nonumber \\
  &= {\rm d}K^\dagger  \wedge K K^\dagger {\rm d}K
  \nonumber \\
  &= {\rm d}K^\dagger K  \wedge K^\dagger {\rm d}K
  \nonumber \\
  &= -(K^\dagger {\rm d}K)  \wedge K^\dagger {\rm d}K
  \nonumber \\
  &= 0 \, ,
  \label{eq:Closed1Form}
\end{align} 
where the penultimate line follows from ${\rm d}(K K^\dagger)=0$
and the last line is a consequence of the antisymmetry of the wedge
product.  Hence, $K^\dagger {\rm d}K$ is a closed 1-form and therefore
on a topologically contractible path is exact.  A pure gauge thus does
not contribute to the NAGP.  As explained in~\cite{Wil84}, for a
general $U(4)$ transformation $G(s)$, it is precisely the (nontrivial)
projection of $G^\dagger {\rm d}G$ in
Eq.~(\ref{eq:GaugePotentialAsMaurerCartan}) onto the subspace
$\mathbb{H}_{(1,1)}$ that can induce a nontrivial gauge field
$\mathcal{A}_{ab}$.  This situation occurs only if nonlocal operations
are used.

\subsection{Decomposition of group elements}
\label{subsec:Decomp}

In order to calculate the NAGP acquired by a state $\rho$ undergoing
cyclic evolution in $U(4)/LO$, we first construct a decomposition of
$U(4)$ into gauge transformations and complementary nonlocal
transformations on the coset space $U(4)/LO$.  Then, with a suitable
parametrization, we derive an expression for the gauge potential
$\mathcal{A}$.  We begin by decomposing the group $U(4)$ in such a way
as to define a simple parametrization for the coset space $U(4)/LO$.

Let $\mathfrak{k} = u(2)^a \times u(2)^b$ be the set of operators
(Hamiltonians) that generate local operations, i.e., the Lie algebra
of $LO$.  A basis for $\mathfrak{k}$ is given by
\begin{align} 
  J_{ax} &= \frac{1}{2} (a_H^\dagger a_V + a_V^\dagger a_H) \, , &\ 
  J_{bx} &= \frac{1}{2} (b_H^\dagger b_V + b_V^\dagger b_H) \, ,
  \nonumber \\
  J_{ay} &= \frac{1}{2{\rm i}} (a_H^\dagger a_V - a_V^\dagger a_H)\, , &
  J_{by} &= \frac{1}{2{\rm i}} (b_H^\dagger b_V - b_V^\dagger b_H)\, , 
  \nonumber \\
  J_{az} &= \frac{1}{2} (a_H^\dagger a_H - a_V^\dagger a_V) \, , &
  J_{bz} &= \frac{1}{2} (b_H^\dagger b_H - b_V^\dagger b_V) \, ,
  \nonumber \\
  J_{a0} &= \frac{1}{2} (a_H^\dagger a_H + a_V^\dagger a_V) \, , &
  J_{b0} &= \frac{1}{2} (b_H^\dagger b_H + b_V^\dagger b_V) \, .
  \label{kGen}
\end{align} 
A complementary set $\mathfrak{p}$ for the Lie algebra $u(4)$ of
$U(4)$ is spanned by the eight elements
\begin{align} 
  J_{HHx} &= \frac{1}{2} (a_H^\dagger b_H + b_H^\dagger a_H) \, , &
  J_{HHy} &= \frac{1}{2{\rm i}} (a_H^\dagger b_H - b_H^\dagger a_H) \, ,
  \nonumber \\ 
  J_{HVx} &= \frac{1}{2} (a_H^\dagger b_V + b_V^\dagger a_H) \, , &
  J_{HVy} &= \frac{1}{2{\rm i}} (a_H^\dagger b_V - b_V^\dagger a_H) \, ,
  \nonumber \\ 
  J_{VHx} &= \frac{1}{2} (a_V^\dagger b_H + b_H^\dagger a_V) \, , &
  J_{VHy} &= \frac{1}{2{\rm i}} (a_V^\dagger b_H - b_H^\dagger a_V) \, ,
  \nonumber \\ 
  J_{VVx} &= \frac{1}{2} (a_V^\dagger b_V + b_V^\dagger a_V) \, , &
  J_{VVy} &= \frac{1}{2{\rm i}} (a_V^\dagger b_V - b_V^\dagger a_V) \, .
  \label{pGen}
\end{align} 
Together, $\mathfrak{k} \oplus \mathfrak{p}$ form a basis for the Lie
algebra of $U(4)$; thus any group element of $U(4)$ can be expressed
as $G = \exp ({\rm i}\sum_i x_i J_i )$ where $x_i$ are real parameters
and the sum is running over all 16 generators $J_i$ of $\mathfrak{k}
\oplus \mathfrak{p}$.

The set $\mathfrak{k}$ is a subalgebra, satisfying
$[\mathfrak{k},\mathfrak{k}] \subset \mathfrak{k}$, and the set
$\mathfrak{p}$ satisfies $[\mathfrak{k},\mathfrak{p}]\subset
\mathfrak{p}$ and $[\mathfrak{p},\mathfrak{p}]\subset \mathfrak{k}$.
These properties enable a Cartan decomposition~\cite{Kna96} of the
group $U(4)$, i.e., any element $G\in U(4)$ can be written in the form
$G= PK$ with $K\in LO$ and $P$ of the form $ \exp ({\rm i}\sum_{\mu}
x_\mu J_\mu )$ where the sum now runs over the set $\mathfrak{p}$
only.

The group element $P$ can be further simplified via the decomposition
\begin{equation} 
  P = \bar{K} P_0 \bar{K}^\dag \, ,
  \label{Pdecomp} 
\end{equation} 
where $\bar{K} \in LO$ and
\begin{equation} 
  P_0 = \exp ({\rm i} x_H J_{HHx}) \exp ({\rm i} x_V J_{VVx})\, .
  \label{P0}
\end{equation} 
A general proof of this decomposition is given in Appendix~\ref{app1}.

Any $\bar K$ can be written by using eight real parameters
and $P_0$ contains two parameters.  Thus, because
$\mathfrak{p}$ is only eight-dimensional, we can further reduce
Eq.~(\ref{Pdecomp}).  Using the Euler
parametrization of $SU(2)$, we express $\bar K$ as
\begin{multline} 
  \bar K = e^{{\rm i} \alpha_a J_{az}}
  e^{{\rm i} \beta_a J_{ay}} e^{{\rm i} \gamma_a J_{az}} e^{{\rm i}
  \delta_a J_{a0}} \\ 
  \times e^{{\rm i} \alpha_b J_{bz}}
  e^{{\rm i} \beta_b J_{by}} e^{{\rm i} \gamma_b J_{bz}} e^{{\rm i}
  \delta_b J_{b0}} \, .
  \label{Eulerdecomp}
\end{multline} 
The isotropy group of $x_V J_{HHx} + x_H J_{VVx}$ is
parametrized by $\alpha_a-\alpha_b$ and $\delta_a + \delta_b$. Thus,
we can define $\alpha_b=\delta_b=0$ in Eq.~(\ref{Pdecomp}).

Thus, we can now express any $G \in U(4)$ in the form
\begin{equation} 
  G = \bar K P_0 K^\prime 
  \label{Gdecomp}
\end{equation} 
with $P_0$ a two-parameter transformation of the form (\ref{P0}),
$\bar K$ a six-parameter transformation of the form
(\ref{Eulerdecomp}) with $\alpha_b=\delta_b=0$, and $K' = \bar{K}^\dag
K \in LO$.  Thus, $K$ is an eight-parameter subgroup describing the
local (gauge) transformations, and $\bar{K} P_0 \bar{K}^{-1}$ is a
complementary eight-parameter set that generates nonlocal
transformations.  This decomposition of group elements is a
generalization of a method applied by Byrd~\cite{Byr98} on $SU(3)$.

\subsection{Gauge potential and Maurer-Cartan form}

Using Eq.~(\ref{AinG}) we are now able to express the gauge potential
in terms of the Maurer-Cartan forms of the group elements
$\bar{K},P_0,K^\prime$.  With the decomposition $G= \bar{K} P_0 K'$ of
Eq.~(\ref{Gdecomp}), we find
\begin{align} 
  \mathcal{A}_{ab} &= {\rm i} \langle \psi_a(0)|G^\dagger {\rm
      d}G|\psi_b(0) \rangle \nonumber \\ 
  &= {\rm i} \langle \psi_a(0) | K^{\prime \dagger} P_0^\dagger
      \bar K^\dagger {\rm d}( \bar K P_0 K') | \psi_b(0) \rangle
      \nonumber \\  
  &= {\rm i} \langle \psi_a(0) | (K^{\prime \dagger } P_0^\dagger)
      (\bar K^\dagger {\rm d}\bar K) (P_0 K^\prime) \nonumber \\ 
  & \qquad
      + K^{\prime \dagger } (P_0^\dagger {\rm d}P_0 ) K^\prime 
      + K^{\prime \dagger } {\rm d}K^\prime | \psi_b(0) \rangle  \\ 
  &= \langle \psi_a(0) | K^{\prime \dagger } P_0^\dagger \Theta_{\bar
      K} P_0 K^\prime \nonumber \\
  &\qquad + K^{\prime \dagger }\Theta_{P_0} K^\prime +
      \Theta_{K^\prime } | \psi_b(0) \rangle   \, .\nonumber
\end{align} 
This expression can be greatly simplified as follows.  The term
containing $\Theta_{K^\prime}$ describes a pure gauge; it therefore
does not contribute to the NAGP as shown by
Eq.~(\ref{eq:Closed1Form}).  The Maurer-Cartan form of $P_0$ can
easily be calculated from Eq.~(\ref{P0}) and is given by
\begin{equation}
  \label{eq:ThetaP0}
  \Theta_{P_0} = -({\rm d}x_H) J_{HHx}- ({\rm d}x_V) J_{VVx} \, .
\end{equation}
This operator, which mixes the components of the two spatial modes $a$
and $b$, maps the subspace $\mathbb{H}_{(1,1)}$ into its complement.
Therefore, because $K^\prime | \psi_b(0) \rangle$ is an element of
$\mathbb{H}_{(1,1)}$, we find that
\begin{equation}
  \label{eq:TermGivesZero}
  \langle \psi_a(0) | K^{\prime \dagger} \Theta_{P_0} K'
  |\psi_b(0)\rangle = 0 \, , \quad \forall\ K' \in LO \, ,
\end{equation}
and thus the contribution of $\Theta_{P_0}$ to the gauge potential is
zero.  As a result, the angles $x_H$ and $x_V$ characterizing the
nonlocal operation $P_0$ only enter the gauge potential as parameters,
and there is no need to integrate over them ($\mathcal{A}$ does not
contain ${\rm d}x_H$ or ${\rm d}x_V$).

The only nontrivial contribution to $\mathcal{A}$ is given by the term
containing $\Theta_{\bar K}$. In this expression, $K^\prime$ only
enters as a gauge transformation, and we can fix the gauge by setting
$K^\prime$ equal to the identity.  We call this choice of gauge the
\emph{entanglement gauge}.  Thus, $G= \bar K P_0$ and the final form
of the gauge potential is
\begin{equation}  
  \mathcal{A}_{ab} = \langle \psi_a(0) | P_0^\dagger \Theta_{\bar K}
  P_0 |\psi_b(0) \rangle \, .
  \label{Afinal}
\end{equation} 
Using the Euler decomposition (\ref{Eulerdecomp}) and with
$\alpha_b=\delta_b=0$, we explicitly find
\begin{multline} 
  \Theta_{\bar K} = -(\cos (\beta_a) {\rm d}\alpha_a +{\rm d}\gamma_a)
  J_{az} -({\rm d}\gamma_b) J_{bz}\\
  -(\cos (\gamma_a)\sin(\beta_a) {\rm d}\alpha_a
  -\sin(\gamma_a) {\rm d}\beta_a) J_{ax}\\
  -(\cos (\gamma_b)\sin(\beta_b) {\rm d}\alpha_b
  -\sin(\gamma_b){\rm d}\beta_b) J_{bx}\\
  -(\sin (\gamma_a)\sin(\beta_a) {\rm d}\alpha_a
  +\cos(\gamma_a){\rm d}\beta_a) J_{ay}\\
  -(\sin (\gamma_b)\sin(\beta_b) {\rm d}\alpha_b +\cos(\gamma_b){\rm
    d}\beta_b) J_{by}\\ - ({\rm d}\delta_a) J_{a0} \, .
  \label{mck}
\end{multline} 
This explicit expression allows us to calculate the gauge potential
$\mathcal{A}_{ab}$ for any path in $U(4)/LO$.

\section{Realizing a non-Abelian geometric phase in quantum interferometry}
\label{sec:NAGP}

With a parametrization of the coset space $U(4)/LO$ and an explicit
expression for the gauge potential in terms of this parametrization,
we can now calculate the NAGP acquired by evolution of a state $\rho$
with support in $\mathbb{H}_{(1,1)}$ about a closed path $\mathcal{C}$
in the geometric space $U(4)/LO$ by a parametrized transformation
$G(s) \in U(4)$.

\subsection{Parametrized transformations}

In order to realize the evolution of a state $\rho$ along a path in
$U(4)/LO$, we must construct an interferometer that evolves $\rho$ as
$G(s)\rho G(s)^\dag$ for some $G(s) \in U(4)$.  Here, $s$ is a
\emph{pseudotime} describing the evolution, and is an adjustable
parameter of the interferometer.

We will realize a closed path $\mathcal{C}$ in the geometric space as
a sequence of transformations in one-parameter subgroups of $U(4)$.
We show in Appendix~\ref{app2} that any one-parameter subgroup $U(s)$
of $U(4)$ can be realized in an optical interferometer using variable
phase shifts and other fixed linear optical elements.  We give
examples below where the number of optical elements is kept very
small.

Let $\{ G_k(s_k); k=1,2,3\}$ be three one-parameter $U(4)$
transformations that perform evolution about a closed path
$\mathcal{C}$, with $s_k \in \mathbb{R}$ the parameter for each path.
The interferometer is constructed to perform the parametrized
operation
\begin{equation}
  \label{eq:TrianglePath}
  G(s_1,s_2,s_3) = G_3(s_3) G_2(s_2) G_1(s_1) \, ,
\end{equation}
which can be used to implement cyclic evolution as follows.
Initially, all parameters are set equal to zero, and the output state
is the initial state $\rho^{(1)}$ in $U(4)/LO$.  Parameter $s_1$ is
made to increase from $0$ to some fixed value $s_1^0$, resulting in
the output state $\rho^{(2)} = G_1(s_1^0) \rho^{(1)} G_1(s_1^0)^{-1}$.
Following this first step, $s_2$ is increased from $0$ to $s_2^0$
yielding the state $\rho^{(3)}$; following that, $s_3$ is increased
from $0$ to $s_3^0$ yielding $\rho^{(4)}$.  The condition for closure
is satisfied if the transformations and parameters are chosen such
that
\begin{equation}
  \label{eq:Closure}
  G_3(s_3^0) G_2(s_2^0) G_1(s_1^0) \in LO \, ,
\end{equation}
and thus $\rho^{(4)} \simeq \rho^{(1)}$.  The cyclic evolution
transports the state about the path $\rho^{(1)} \to \rho^{(2)} \to
\rho^{(3)} \to \rho^{(4)}$, where $\rho^{(4)} \simeq \rho^{(1)}$ are
both states with support in $\mathbb{H}_{(1,1)}$.

Using our parametrization of elements in $U(4)$ as given in
Sec.~\ref{subsec:Decomp}, we explicitly calculate the NAGP acquired
via evolution about such a path.  To identify closed paths, we note
that $P_0(x_H,x_V)$ of Eq.~(\ref{P0}) only maps $\mathbb{H}_{(1,1)}$
onto itself if $x_H = m\pi$ and $x_V = n\pi$, for $m,n$ integers that
are either both even or both odd.  These two conditions identify
endpoints for a closed path on $U(4)/LO$.

We now give an explicit construction for evolution about a closed
path.  Define $G_1(s_1)$ to be
\begin{align}
  \label{eq:U1}
  G_1(s_1) &= \exp\bigl({\rm i} s_1 (\cos^2 \theta J_{HHx} + \sin^2 \theta
  J_{VVx})\bigr) \nonumber \\ 
  &= P_0(s_1 \cos^2 \theta, s_1 \sin^2 \theta) \, ,
\end{align}
where $\theta \in [0,\pi/2]$.  This transformation, realized using two
parametrized polarizing beamsplitters, is inherently nonlocal and
evolves any state with support in $\mathbb{H}_{(1,1)}$ to a state with
support not entirely within this subspace.  The second transformation
$G_2(s_2)$ is defined to be
\begin{equation}
  \label{eq:U2}
  G_2(s_2) = \bar K(s_2) \, ,
\end{equation}
where $\bar K(s_2)$ is an arbitrary one-parameter subgroup of $LO$
such that $\bar K(0)$ is the identity.  This transformation is
implemented using polarization rotators and phase shifters in each
mode.  Finally, the transformation $G_3(s_3)$ is defined to be
\begin{equation}
  \label{eq:U3}
  G_3(s_3) = \bar K P_0(s_3 \cos^2 \phi, s_3 \sin^2 \phi) \bar K^\dag \, ,
\end{equation}
with $\bar K = \bar K(s_2^0)$, $\phi \in [0,\pi/2]$ and $s_3^0$ is
chosen such that
\begin{gather}
  \label{eq:ClosureCondition}
  s_1^0 \cos^2 \theta + s_3^0 \cos^2 \phi = m\pi \, , \\ 
  s_1^0 \sin^2 \theta + s_3^0 \sin^2 \phi = n\pi \, ,
\end{gather}
for some integers $m,n$ (either both even or both odd).  This final
transformation is implemented using a combination of polarization
rotators and phase shifters (to realize $\bar K$ and $\bar K^{-1}$)
along with two parametrized polarizing beamsplitters (to realize
$P_0$).  With these conditions, the $G_k(s_k)$ satisfy
\begin{equation}
  \label{eq:Closed}
  G_3(s_3^0) G_2(s_2^0) G_1(s_1^0) = \bar K P_0(m\pi,n\pi) \in LO \, .
\end{equation}
 
The gauge potential is zero on paths 1 and 3 because $\mathcal{A}$ does
not contain ${\rm d}x_H$ or ${\rm d}x_V$ and all other differentials
are zero.  The only contribution to the NAGP therefore comes from path
2 on which we find
\begin{equation}
  \label{eq:ExplicitGaugePotential} 
  \mathcal{A}_{ab} = \langle \psi_a(0) | P^\dagger \Theta_K P
  |\psi_b(0) \rangle \, ,
\end{equation}
where $P = P_0(s_1^0 \cos^2 \theta, s_1^0 \sin^2 \theta)$.  Thus, for
our calculations, we only require $P_0$ and can otherwise ignore the
evolution on paths 1 and 3.  

Thus, one can use a $U(4)$ interferometer to evolve a state $\rho$
about a closed path in $U(4)/LO$, and calculate the acquired NAGP by
Eqs.~(\ref{eq:WilsonLoop}) and (\ref{eq:ExplicitGaugePotential}).  The
net transformation on the state $\rho$ upon cyclic evolution will
consist of a NAGP due to the geometry of the geometric space and a
local gauge transformation; thus, it is not possible to measure the
NAGP directly.  The effects of this geometric phase can be seen,
however, by varying the cyclic paths used.  In the following we give a
specific example of this procedure.

\subsection{An Example}
\label{subsec:Example}

As an explicit example, we choose $G_1(s_1)$ according to
Eq.~(\ref{eq:U1}) by setting $\theta = \pi/4$ and $s_1^0 = \pi$.
Thus,
\begin{equation}
  \label{eq:Example1G1}
  G_1(s_1) = \exp({\rm i} s_1( J_{HHx}+J_{VVx})/2) \, , 
\end{equation}
which corresponds to a (polarisation-independent) beamsplitter with
reflectivity $r = \sin^2 s_1$.  As discussed above, this parametrized
transformation can be constructed using a variable phaseshifter
(parametrized by $s$) and fixed linear optics, in this case, in the
form of a Mach-Zehnder interferometer.

\begin{figure}
  \includegraphics*[width=3.25in,keepaspectratio]{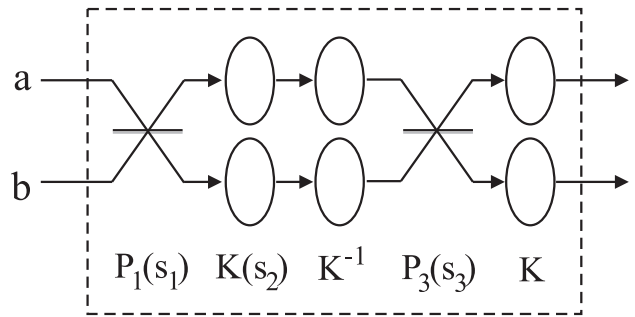}
  \caption{Diagram of an interferometer that would realize a cyclic
    evolution exhibiting a NAGP.  The two spatial modes, $a$ and $b$,
    have inputs on the left.  These spatial modes are combined twice
    at parametrized beamsplitters, realizing the nonlocal operations
    $P_1(s_1) = P_0(s_1/2,s_1/2)$ and $P_3(s_3) = P_0(s_3/2,s_3/2)$.
    Local operations are represented as circles in one spatial mode
    only, realizing the parametrized operation $K(s_2)$ and the fixed
    operations $K$, $K^{-1}$.  Note that the cyclic evolution does not
    occur as the state passes through the interferometer \emph{in
      time}; instead, the evolution occurs in pseudotime as the
    parameters $s_i$ are increased from $0$ to $s_i^0$ in sequence.}
  \label{fig:Interf}
\end{figure}

For the second portion of the transformation, we choose $G_2(s_2) =
\bar{K}(s_2)$ according to the parametrization of
Eq.~(\ref{Eulerdecomp}) with $s_2 = 2\beta_a = 2\beta_b$ and all other
parameters set to zero.  Thus,
\begin{equation}
  \label{eq:Example1G1}
  G_2(s_2) = \exp({\rm i} s_2( J_{ay}+J_{by})/2) \, , 
\end{equation}
with some arbitrary endpoint $s_2^0 > 0$.  This transformation can be
implemented using polarization rotation and phase shifts in each arm
(local operations).  Finally, to complete the cyclic evolution, we
choose $G_3$ according to Eq.~(\ref{eq:U3}) with $\phi = \pi/4$ and
$s_3^0 = \pi$ in order to satisfy the closure condition
(\ref{eq:ClosureCondition}); again, this transformation can be
performed using a combination of local operations and a
polarization-independent beamsplitter.  The interferometer that would
realize this cyclic evolution is depicted in Fig.~\ref{fig:Interf}.

Again, we note that only the second path contributes to the NAGP,
which we now calculate explicitly.  The Maurer-Cartan form for the
transformation $\bar{K}(s_2)$ is
\begin{equation}
  \label{eq:Example1MaurerCartan}
  \Theta_{\bar{K}} = - \frac{{\rm d}s_2}{2} (J_{ay}+J_{by}) \, .
\end{equation}
Using Eq.~(\ref{eq:ExplicitGaugePotential}) and the basis states for
$\mathbb{H}_{(1,1)}$,
\begin{align}
  \label{eq:BasisStates}
  |\psi_1(0)\rangle &= |HH\rangle \, , &
  |\psi_2(0)\rangle &= |HV\rangle \, , \nonumber\\
  |\psi_3(0)\rangle &= |VH\rangle \, , &
  |\psi_4(0)\rangle &= |VV\rangle \, ,
\end{align}
we calculate the gauge potential to be
\begin{equation}
  \label{eq:Example1GaugePotential}
  \mathcal{A}_{ab} = \frac{{\rm d}s_2}{4i} 
  \begin{pmatrix} 0 & -1 & -1 & 0 \\ 
                  1 & 0 & 0 & -1 \\
                  1 & 0 & 0 & -1 \\
                  0 & 1 & 1 & 0 \end{pmatrix} \, .
\end{equation}
Integrating along the path gives
\begin{equation}
  \label{eq:Example1NAGP} 
  K_{\rm NAGP} = \exp({\rm i} \int_0^{s_2^0} \mathcal{A}) =
  \begin{pmatrix}
             C_+ & -S  & -S  & C_- \\ 
             S   & C_+ &-C_- & -S  \\ 
             S   &-C_- & C_+ & -S  \\ 
             C_- & S   &  S  & C_+ \\ 
           \end{pmatrix}  
\end{equation} 
with $S= \frac{1}{2}\sin (s_2^0/2)$ and $C_\pm = \frac{1}{2}(1\pm
\cos(s_2^0/2))$.

With such a setup, it is possible to transform states in
$\mathbb{H}_{(1,1)}$ about closed paths, and observe the effects of
the geometric phase.  The total transformation on the state will be
$\bar{K}(s_2^0)P_0(\pi,\pi)$, which consists of a combination of a
NAGP given by Eq.~(\ref{eq:Example1NAGP}) and a local gauge
transformation.  The effect of the NAGP can be isolated and observed
by varying $s_2^0$ and evolving a given state about many different
paths.  In this way, the NAGP can be shown to depend on the geometry
of the system, i.e., on the choice of closed path $\mathcal{C}$.

We note that the quantum tomographic techniques of White \emph{et
  al}~\cite{Whi02} have demonstrated that a two-photon state can be
completely characterized (with sufficiently many copies of the state
or iterations of the experiment).  These techniques can be employed in
the proposed setup described above to measure the NAGP.  Note again
that the non-Abelian ``phase'' described here is a U(4) transformation
on a bi-partite density matrix, and thus its effect can readily be
observed on an appropriate set of initial states.

It is also possible to realize other cyclic evolutions by choosing
different interferometric setups.  Another example would be to choose
$\theta=0$ and $s_1^0 = \pi$ together with $\phi=0$ and $s_3^0=\pi$;
this choice corresponds to exploying variable beamsplitters which mix
only the horizontal (vertical) components for transformations $G_1$
($G_3$), respectively.  In general, the resulting NAGPs acquired by
cyclic evolutions about these different paths will not commute; this
non-commutative property is what distinguishes the NAGP from its
Abelian counterparts.

\section{Discussion}
\label{sec:Conc}

Gauge symmetry has proven to be immensely important in theoretical
physics, and we have established both a language and a method for
applying gauge theory to quantum information theory.  The entanglement
gauge introduced here establishes an equivalence between actions that
differ only by local operations.  Nonlocal properties, such as the
crucial resource of entanglement, are then regarded as quantities that
are invariant under entanglement gauge transformations.  Although our
focus has been on developing an entanglement gauge for actions on
qubits and on pairs of qubits, this analysis could be extended to
coupled qudit-qudit systems~\cite{Bar02}, but of course the group
$U(4)$ would need to be replaced by the appropriate larger group.

Gauge theory enables the dynamics to be interpreted geometrically. One
way to access information about the geometry by experimental means is
via measurements of the geometric phase, and we apply entanglement
gauge theory to photonic qubits in interferometry as an experimental
means for manifesting and measuring a non-Abelian geometric phase.
Although geometric phase experiments for systems with non-Abelian
dynamics have been proposed~\cite{San01}, we have proposed here the
first controlled experiment for realizing a \emph{non-Abelian
  geometric phase} as opposed to an Abelian geometric phase for a
system with non-Abelian dynamics.

It is particularly interesting that this non-Abelian geometric phase
is manifested in interferometry, which involves manipulations of the
electromagnetic field as the electromagnetic field exhibits a $U(1)$
gauge symmetry.  The essential point here in realizing a non-Abelian
geometric phase is that the gauge symmetry arises through the
equivalence under local operations. We consider sources of entangled
qubits, and transform the state to one that is equivalent under local
operations with an accumulated non-Abelian geometric phase.

The transformation of the state via linear optics and including an
accumulation of a non-Abelian geometric phase requires the state,
during its evolution, to leave the space consisting of a single photon
in each of the two channels of the interferometer.  In leaving this
space the evolution includes support from having two photons in one
channel, increasing the dimension of the Hilbert space from four to
the full ten dimensions of the two-photon irrep of $U(4)$.  This
evolution out of the four-dimensional Hilbert space is not problematic
with respect to the entanglement gauge and the resultant geometry,
though, because the interferometer functions as a `black box', as we
have carefully described. The key concept here is that the output
state is controlled by parameters of the interferometer (which we
think of as being controlled by knobs), and the output state evolves
along a path in the geometric space as a function of the pseudotime
determined by the settings of these knobs.  The output state can then
be evolved (with evolution parametrized by pseudotime) up to a state
equivalent to the initial state up to local operations, with an
acquired non-Abelian geometric phase.

We also note that it is possible to map a large class of pseudotime
evolutions to a corresponding propagation in real time. This map can
be achieved by considering the optical elements (e.g., a phase
shifter) as implementing continuous unitary transformations as the
photon wavepackets are propagating through the elements.  In this
sense an optical element that induces a change in pseudotime from
$s_0$ to $s_1$ would correspond to a continuous real-time evolution
from $t(s_0)$ to $t(s_1)$.  By treating each optical element in this
way, one arrives at a real- instead of a pseudotime evolution.

For propagation in real time, it is necessary to consider the
\emph{dynamical} evolution of the subspace $\mathbb{H}_{(1,1)}$
introduced in Sec.~\ref{sec:Interferometry}.  Because the dynamical
evolution operator $U_D := \exp (-i H t/\hbar)$ associated with a
Hamiltonian $H$ may not commute with the NAGP of
Eq.~(\ref{eq:WilsonLoop}), the NAGP becomes difficult to
isolate~\cite{Ana88}.  One of the advantages of using optical devices
is that this problem can be circumvented; the two photons possess the
same energy and the Hamiltonian on the Hilbert space $\mathbb{H}_2$ is
proportional to the identity operator.  Thus, the dynamical evolution
$U_D$ always commutes with the NAGP and is not relevant.

We have presented the mathematical tools for designing photonic qubit
experiments and for determining the resultant non-Abelian geometric
phase from a closed-path evolution. A specific example has also been
provided in Sec.~\ref{sec:NAGP} to provide a clear illustration about
how such an experiment would be conducted. We observe that three
components in the `black box' interferometer are controlled by the
pseudotime parameter $s$, and the output state can then be measured
by tomographic means: recent tomographic experiments~\cite{Whi02} have
in fact demonstrated the feasibility of such measurements.  Generating
pure states of entangled photonic qubits, transforming such states via
interferometry and measuring the output states via tomography are thus
all feasible technologies.  Thus, the experimental manifestations of
the entanglement gauge are within the reach of current technology.

In summary we have introduced the entanglement gauge and developed the
non-Abelian geometric phase as an experimentally realizable
manifestation of the entanglement gauge. With the rapid growth in
quantum information theory, the entanglement gauge provides a new
approach to tackling issues such as analyzing equivalence under local
operations, realizing geometric phases in quantum information
experiments and connecting operations used in quantum information to
principles of differential geometry.  Recent proposed applications of
geometric phases to fault-tolerant quantum
computation~\cite{Jon00,Zan99} suggests that such phases may be useful
for quantum information processing.  We trust that our entanglement
gauge formalism presented here will also prove to be a useful tool for
investigations into applications of entanglement, such as entanglement
distillation~\cite{Ben96}, distributed quantum
computation~\cite{Eis00,Cir99}, and the ability to perform nonlocal
operations using entanglement, local operations and classical
communication~\cite{Eis00,Col01,Dur01}.

\appendix

\section{Proof of decomposition}
\label{app1}

The proof of Eq.~(\ref{P0}) can be given in a more general form which
also applies to a large degree to $U(n)$ in general. To do so we start
with a general matrix $M$ of dimension $p\times q$ and consider the
exponentation of the $(p+q)\times (p+q)$ matrix,
\begin{equation} 
  \exp \left [ i \left ( \begin{array}{cc}
            0 & M \\ M^\dagger  & 0
           \end{array} \right ) \right ]
\label{expo} \end{equation} 
One can prove by induction that
\begin{equation} 
   \left ( \begin{array}{cc}
            0 & M \\ M^\dagger  & 0
           \end{array} \right )^{2n} =
   \left ( \begin{array}{cc}
            (MM^\dagger )^n & 0  \\ 0 & (M^\dagger M)^n
           \end{array} \right ) 
\end{equation}  
which can be used to give a convenient series expansion of the
exponential. One now can exploit the singular value decomposition $M =
U D U^{\prime \dagger }$, where $U$ and $U^\prime $ are unitary
matrices of dimension $p\times r$ and $q\times r$, and $r$ the minimum
of $p$ and $q$. The matrix $D$ is a real diagonal matrix of dimension
$r\times r$.  It is then easy to show that $(MM^\dagger )^n = U D^{2n}
U^\dagger $ as well as $(M^\dagger M)^n = U^\prime D^{2n} U^{\prime
  \dagger} $.  Consequently, the exponential can be written as
\begin{multline} 
  \exp \left [ i \left ( \begin{array}{cc}
            0 & M \\ M^\dagger  & 0
           \end{array} \right ) \right ] \\ =
  \left ( \begin{array}{cc}
           U  & 0\\ 0& U^\prime 
           \end{array} \right ) 
  \left ( \begin{array}{cc}
           \cos D  & i \sin D \\ i \sin D & \cos D
           \end{array} \right ) 
  \left ( \begin{array}{cc}
           U^\dagger   & 0 \\ 0 & U^{\prime \dagger }
           \end{array} \right ) 
  \label{decomp1}
\end{multline} 

Restricting to the case $p=q=2$ the exponential corresponds to a
general element $P$ in the adjoint representation.  In our notation
this representation corresponds to the four-dimensional one-photon
representation of the generators. The matrices $U$ and $U^\prime$ are
then general $U(2)$ transformations and just describe the local
operations of the subgroup ${\cal K}$. Since $D$ then just contains
two independent parameters we find that any nonlocal operation $P$ can
be composed of a local operation and a certain nonlocal operation that
depends on two parameters only. This result is not in conflict with
that of Refs. \cite{Cir91,mm90} in which three independent parameters
are found since in these papers the local operations are restricted to
$SU(2) \otimes SU(2)$ transformations only. In our approach we also
consider the relative phase shift generated by $J_{a0} -J_{b0}$, which
is not an element of $SU(2) \otimes SU(2)$ but of $U(2) \otimes U(2)$,
as a local operation.

The explicit form of the middle matrix of the r.h.s. of
Eq.~(\ref{decomp1}) just corresponds to the form of $P_0$ with
$\gamma_i$ equal to the diagonal entries of $D$.  As this result holds
in the adjoint representation and (\ref{decomp1}) is
representation-independent, we infer that the result holds for any
representation.  This concludes the proof.  We remark that this result
has a straightforward extension to an $n+n$ decomposition of $U(2n)$.

\section{Constructing parametrized $U(4)$ operations}
\label{app2}

In order to realize a NAGP, it is necessary to perform parametrized
$U(4)$ transformations using an optical interferometer.  In this
appendix, we prove that any one-parameter subgroup $G(s)$ in $U(4)$
can be constructed out of variable phase shifts in each mode and fixed
optical elements (such as beamsplitters).  This result is a
generalization of the Mach-Zehnder inferferometer, where any $SU(2)$
transformation can be implemented using fixed-reflectivity
beamsplitters and a variable phase shift in one arm.

A phase shift of one mode (with annihilation operator $c$) is
described as a $U(1)$ transformation, generated by an operator of the
form $c^\dag c$.  Thus, for a $U(4)$ interferometer with four modes,
phase shifts in each mode form a subgroup $U(1) \times U(1) \times
U(1) \times U(1) \subset U(4)$.  Consider a one-parameter (variable)
phase shift for the four modes of the $U(4)$ interferometer, generated
by an operator of the form
\begin{equation}
  \label{eq:VarPhaseShiftGen}
  D = c_1 a_H^\dag a_H + c_2 a_V^\dag a_V +c_3 b_H^\dag b_H +c_4
  b_V^\dag b_V \, ,
\end{equation}
for some real coefficients $c_i$.  The parametrized transformation
\begin{equation}
  \label{eq:VarPhaseShift}
  S(s) = \exp({\rm i}s D) \, ,
\end{equation}
thus describes a one-parameter variable phase shift in the four modes,
where the relative phase shifts between each of the modes are
determined by the coefficients $c_i$.

For an arbitrary one-parameter subgroup $G(s)$ of $U(4)$, there exists
a fixed matrix $V$ that diagonalizes $G(s)$ for all $s$; i.e.,
\begin{equation}
  \label{eq:DiagonalizeG}
  V G(s) V^{-1} = S(s) \, ,
\end{equation}
for some $S(s)$ of the form (\ref{eq:VarPhaseShift}).  Thus, with the
ability to implement the variable phase shift transformation $S(s)$
and the fixed transformation $V$ (and thus also $V^{-1}$), the
one-parameter subgroup $G(s)$ can be implemented as $G(s) = V^{-1}
S(s) V$.

In addition, it has been shown in~\cite{deG01} than any (fixed) element
in $SU(n)$ can be factorized into a product of $SU(2)$
transformations.  With this result, it is possible to construct the
required fixed transformations $V$ and $V^{-1}$ out of beamsplitters,
phase shifters, and polarization rotations.

\begin{acknowledgments}
  We acknowledge helpful discussions with H.\ de Guise, D.\ J.\ Rowe
  and A.\ G.\ White.  This project has been supported by Macquarie
  University, the Australian Research Council, the Optikzentrum
  Konstanz and the Forschergruppe Quantengase.
\end{acknowledgments}


\begin{thebibliography}{99}
\bibitem{Bel64} J.\ Bell, Physics \textbf{1}, 195 (1964);
        Rev.\ Mod.\ Phys.\ \textbf{38}, 447 (1966).
        
\bibitem{Ben93} C.\ H.\ Bennett, G.\ Brassard, C.\ Cr\'{e}peau,
        R.\ Jozsa, A.\ Peres and W.\ K.\ Wootters, \prl \textbf{70},
        1895 (1993).
        
\bibitem{Eke91} A. K. Ekert, \prl \textbf{67}, 661 (1991).
  
\bibitem{Cir97} I.\ Cirac and N.\ Gisin, Phys.\ Lett.\ A \textbf{229},
  1 (1997).
  
\bibitem{Fuc97} C.\ A.\ Fuchs, N.\ Gisin, R.\ B.\ Griffiths, C.-S.\ 
  Niu and A.\ Peres, \pra \textbf{56}, 1163 (1997).

\bibitem{Bra99} S. L. Braunstein, C. M. Caves, R. Jozsa, N. Linden,
  S. Popescu and R. Schack, \prl \textbf{83}, 1054 (1999).

\bibitem{Eke98} A. Ekert and R. Jozsa, Philos. Trans. R. Soc., London
  Ser. A \textbf{356}, 1769 (1998).

\bibitem{Ben92} C. H. Bennett and S. J. Wiesner, \prl \textbf{69}, 2881
  (1992).

\bibitem{Eis00} J. Eisert, K. Jacobs, P. Papadopoulos and
  M. B. Plenio, \pra \textbf{62}, 052317 (2000).

\bibitem{Cir99} J. I. Cirac, A. K. Ekert, S. F. Huelga and
  C. Macchiavello, \pra \textbf{59}, 4249 (1999).

\bibitem{Ben96} C. H. Bennett, G. Brassard, S. Popescu, B. Schumacher,
  J. A. Smolin, and W. K. Wootters, \prl \textbf{76}, 722 (1996).

\bibitem{Egu80} T. Eguchi, P.B. Gilkey and A.J. Hanson, 
  Phys. Rep. {\bf 66}, 213 (1980).

\bibitem{Ber84} M.\ V.\ Berry, Proc.\ Roy.\ Soc.\ (Lond.) \textbf{392},
        45 (1984).
        
\bibitem{Sim83} B. Simon, \prl \textbf{51}, 2167 (1983); F. 
        Wilczek and A. Shapere, \textit{Geometric Phases in Physics},
        Advanced Series in Mathematical Physics - Vol. \textbf{5} (World
        Scientific, Singapore, 1989).
        
\bibitem{Fue02} I. Fuentes-Guridi, A. Carollo, S. Bose and V.
        Vedral, \prl \textbf{89}, 220404 (2002); I. Fuentes-Guridi, S.
        Bose and V. Vedral, \prl \textbf{85}, 5018 (2000).

\bibitem{Jon00} J.\ A.\ Jones, V.\ Vedral, A.\ Ekert, and G.\ 
        Castagnoli, Nature \textbf{403}, 869 (2000).
        
\bibitem{Zan99} P. Zanardi and M. Rasetti, Phys.~Lett.~A
        \textbf{264}, 94 (1999); J. Pachos, P. Zanardi and M. Rasetti,
        \pra \textbf{61}, 010305 (2000).

\bibitem{Kwi95} P.\ G.\ Kwiat, K.\ Mattle, H.\ Weinfurter, A.\
        Zeilinger, A.\ V.\ Sergienko, and Y.\ Shih,
        \prl \textbf{75}, 4337 (1995).
        
\bibitem{Whi02} A. G. White, D. F. V. James, W. Munro and P. 
        G. Kwiat, \pra \textbf{65}, 012301 (2002).
        
\bibitem{Kwi91} P.\ G.\ Kwiat and R.\ Y.\ Chiao, \prl
        \textbf{66}, 588 (1991).

\bibitem{Sjo00a} E.\ Sj\"oqvist, \pra \textbf{62}, 022109
        (2000); B.\ Hessmo and E.\ Sj\"oqvist, \pra \textbf{62},
        062301 (2000).
        
\bibitem{Sjo00b} E.\ Sj\"oqvist, A.\ K.\ Pati, A.\ Ekert, J.\ S.\ 
        Anandan, M.\ Ericsson, D.\ K.\ L.\ Oi, and V.\ Vedral, \prl
        \textbf{85}, 2845 (2000).
        
\bibitem{Cir95} J. I. Cirac and P. Zoller, \prl \textbf{74},
        4091 (1995).
        
\bibitem{Bar01} S. D. Bartlett, D. A. Rice, B. C. Sanders, J. Daboul,
        and H. de Guise, \pra \textbf{63}, 042310 (2001).

\bibitem{Lut99} N. L\"utkenhaus, J. Calsamiglia and K.-A. Suominen,
    \pra \textbf{59}, 3295 (1999).
    
\bibitem{Chu95} I. L. Chuang and Y. Yamamoto, \pra \textbf{52}, 3489
    (1995).
  
\bibitem{San01} B. C. Sanders, H. de Guise, S. D. Bartlett and
        W. Zhang, \prl \textbf{86}, 369 (2001).
        
\bibitem{deG01} H. de Guise, B. C. Sanders, S. D. Bartlett and
  W. Zhang, Czech. J. Phys. \textbf{51}, 312 (2001).

\bibitem{Wil84} F. Wilczek and A. Zee, \prl \textbf{52},
        2111 (1984).
  
\bibitem{Ana88} J.\ Anandan, Phys.\ Lett.\ A \textbf{133}, 171 (1988).
  
\bibitem{Kna96} A.W. Knapp, \textit{Lie groups beyond an introduction}
  (Birkhauser, Boston, 1996).

\bibitem{Byr98} M. Byrd, J.~Math.~Phys.~\textbf{39}, 6125 (1998);
  Erratum ibid \textbf{41}, 1026 (2000).

\bibitem{Bar02} S. D. Bartlett, H. de Guise and B. C. Sanders, \pra
  \textbf{65}, 052316 (2002).
  
\bibitem{Col01} D. Collins, N. Linden and S. Popescu, \pra \textbf{64},
  032302 (2001).
  
\bibitem{Dur01} W. D\"ur and J. I. Cirac, \pra \textbf{64}, 012317
  (2001).

\bibitem{Cir91} W. D\"ur, G. Vidal, J.I. Cirac, N. Linden, and S.
  Popescu, \prl \textbf{87}, 137901 (2001).

\bibitem{mm90} Y. Makhlin, e-print quant-ph/0002045.

\end{thebibliography}
\end{document}